\documentclass[prl,twocolumn,superscriptaddress,showpacs,preprintnumbers,amsmath,amssymb]{revtex4}

\usepackage{graphicx}
\usepackage{dcolumn}
\usepackage{bm}

\usepackage{color}

\begin{document}

\newcommand{\Gr}{\mbox{$\Gamma_1$}}
\newcommand{\GdE}{\mbox{$\Gamma_{2E}$}}
\newcommand{\GdF}{\mbox{$\Gamma_{2F}$}}
\newcommand{\GpEexp}{\mbox{$\Gamma_{\! \varphi\! E}$}}
\newcommand{\GpFexp}{\mbox{$\Gamma_{\! \varphi\! F}$}}
\newcommand{\GpE}{\mbox{$\Gamma_{\! \varphi\! E}^g$}}
\newcommand{\GpF}{\mbox{$\Gamma_{\! \varphi\! F}^g$}}
\newcommand{\eiphiE}{\mbox{$\langle e^{i\varphi (t)} \rangle_E$}}

\newcommand{\tr}{\color{red}}
\newcommand{\tb}{\color{black}}

\title{Decoherence of flux qubits due to {\boldmath $1/f$} flux noise}

\author{F. Yoshihara}
\affiliation{The Institute of Physical and Chemical Research (RIKEN), Wako, Saitama 351-0198, Japan}

\author{K. Harrabi}%
\affiliation{CREST-JST, Kawaguchi, Saitama 332-0012, Japan}

\author{A. O. Niskanen}
\affiliation{CREST-JST, Kawaguchi, Saitama 332-0012, Japan}
\affiliation{VTT Technical Research Centre of Finland, Sensors, P.O. Box 1000, 02044 VTT, Finland}

\author{Y. Nakamura}
\affiliation{The Institute of Physical and Chemical Research (RIKEN), Wako, Saitama 351-0198, Japan} \affiliation{CREST-JST, Kawaguchi, Saitama 332-0012, Japan}
\affiliation{NEC Fundamental and Environmental Research Laboratories, Tsukuba, Ibaraki 305-8501, Japan}

\author{J. S. Tsai}
\affiliation{The Institute of Physical and Chemical Research (RIKEN), Wako, Saitama 351-0198, Japan} \affiliation{CREST-JST, Kawaguchi, Saitama 332-0012, Japan}
\affiliation{NEC Fundamental and Environmental Research Laboratories, Tsukuba, Ibaraki 305-8501, Japan}

\date{\today}

\begin{abstract}
We have investigated decoherence in Josephson-junction flux qubits.
Based on the measurements of decoherence at various bias conditions, we discriminate contributions of different noise sources.
We present a Gaussian decay function extracted from the echo signal as evidence of dephasing due to $1/f$ flux noise whose spectral density is evaluated to be about $(10^{-6} \Phi_0)^2$/Hz at 1~Hz.
We also demonstrate that at an optimal bias condition where the noise sources are well decoupled the coherence observed in the echo measurement is mainly limited by energy relaxation of the qubit.
\end{abstract}


\maketitle


A Josephson-junction qubit is a macroscopic quantum object realized in superconducting electric circuits \cite{Makhlin01}.
As a solid-state device sitting in the midst of the circuit and on top of the substrate, the qubit inherently interacts with various degrees of freedom which decohere the qubit.
For understanding and suppressing the decoherence mechanisms, it is important to identify the dominant noise sources among the many possibilities.

A strategy to tackle this problem is to use the tunability of the qubit.
By changing a bias parameter, one can control the qubit energy as well as the coupling between the qubit and a particular degree of freedom in the environment \cite{Vion02}.
Therefore, the measurement of the decoherence as a function of the bias parameters provides valuable information about the environmental degrees of freedom.
Such a strategy has been applied to charge-phase qubits \cite{Cottet02,Ithier05}, charge qubits \cite{Astafiev04,Duty04}, flux qubits \cite{Bertet05}, and phase qubits \cite{Simmonds04,Claudon06}.

In this Letter, we report a detailed characterization of decoherence in flux qubits based on a similar approach.
In a range of bias conditions, we identify flux noise as the dominant source of the qubit dephasing.
We observe dephasing due to $1/f$ flux noise which gives rise to a characteristic Gaussian decay function that seems to be hidden in the previous studies by other decoherence mechanisms  \cite{Ithier05,Bertet05}.
We also quantify the noise spectrum density.
On the other hand, we show that at an optimal bias condition, where those noises are effectively decoupled from the qubit, the echo decay rate ($\sim$ (3.5~$\mu$s)$^{-1}$ in one sample) is about a half of the energy relaxation rate, and is mainly limited by the latter process.

\begin{figure}[b]
\includegraphics[width=\linewidth]{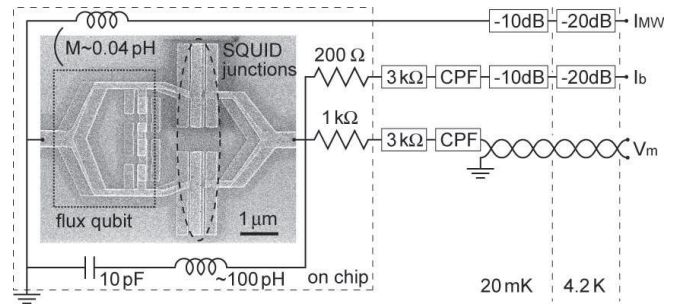}
\caption{
Scanning electron micrograph of a sample and a sketch of the measurement setup.
The qubit shares a part of the loop with a SQUID shunted by an on-chip capacitor.
A current bias line (${\rm I_b}$) and a voltage measurement line (${\rm V_m}$) consist of lossy coaxial cables and are connected to the SQUID via on-chip resistors.
Microwave current pulses are fed through an on-chip control line (${\rm I_{MW}}$) inductively coupled to the qubit.
The sample chip is enclosed in a copper shield box, and all the wires are electrically shielded as well.
CPF stands for a copper-powder filter.
Magnetic flux bias is applied with an external superconducting coil connected to a battery-powered current source.
The sample is cooled to 20~mK in a dilution refrigerator magnetically shielded with three $\mu$-metal layers at room temperature.
}
\label{SEMetc}
\end{figure}

The effective Hamiltonian of a flux qubit can be written as $ H_0 = -\frac{\varepsilon}{2} \sigma_z - \frac{\Delta}{2} \sigma_x $, where $\sigma_z$, $\sigma_x$ are Pauli matrices \cite{Mooij99}.
The energy difference between the two eigenstates is given by $E_{01} = \sqrt{\varepsilon ^2 + \Delta ^2}$.
Here $\Delta$ is the tunnel splitting between two states with opposite directions of persistent current along the loop, $\varepsilon = 2 I_p \Phi_0 (n_\phi - n_\phi^*)$ is the energy bias between the two states, and $I_p$ is the persistent current along the qubit loop.
We define the normalized magnetic flux in the qubit loop $n_\phi \equiv \Phi_{\rm ex}/\Phi_0$, using the externally applied flux through the qubit $\Phi_{\rm ex}$ and the flux quantum $\Phi_0$.
As the flux qubit we study is coupled to a SQUID for readout (Fig.~\ref{SEMetc}), the SQUID bias current $I_b$ also affects flux in the qubit.
The normalized $I_b$-dependent flux is thus defined as $n_\phi^*(I_b) \equiv 0.5 - \Phi_{I_b}/\Phi_0$ so that $n_\phi=n_\phi^*$ corresponds to $\varepsilon =0$, i.e., degeneracy of the two persistent current states.
Note that $I_b$ affects the qubit only through the flux $\Phi_{I_b}(I_b)$ which is induced by the circulating current in the SQUID loop \cite{commentPhi}.

We characterize the coherence of the qubit through measurements of energy relaxation, free-induction decay (FID), and echo decay.
From those measurements, we deduce the energy-relaxation rate as well as the pure dephasing components in the decoherence.
Pure dephasing is caused by the fluctuations of $E_{01}$ which are induced by the fluctuations of parameters $\lambda_i$ in the Hamiltonian.
Here $\lambda_i$'s could be $n_\phi$ and $I_b$ as well as other implicit parameters inside $\Delta$ and $I_p$, such as the island charge offsets and the junction critical currents in the qubit.
Through the two externally controlled parameters $n_\phi$ and $I_b$, we vary the coupling factor $\frac{\partial E_{01}}{\partial \lambda_i}$ and observe the change in the dephasing.
Since the derivatives $\frac{\partial E_{01}}{\partial \lambda_i}$ depend differently from each other on the control parameters $n_\phi$ and $I_b$, and assuming that the fluctuations of $\lambda_i$ are independent of the control parameters, we can separate the contributions in the dephasing from different sources.

\begin{figure}[t]
\includegraphics[width=65mm]{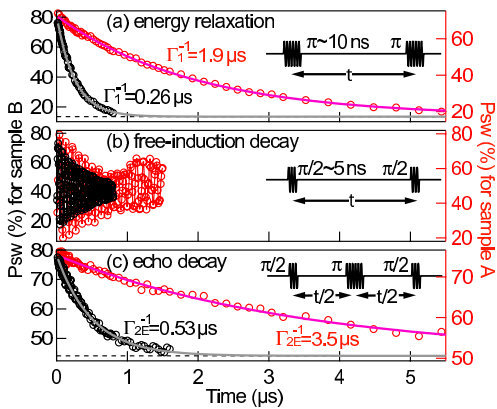}
\caption{
(color online).
Measurements of (a) energy relaxation, (b) free-induction decay, and (c) echo decay at the optimal bias condition $n_\phi=n_\phi^*$ and $I_b=I_b^*$.
Red plots are for sample~A, and black ones are for sample~B\@.
The insets depict the applied pulse sequence for each measurement.
In the measurement of free-induction decay, the microwave carrier frequency was detuned from the qubit frequency $E_{01}/h$ by 10~MHz.
Solid curves in (a) and (c) are exponential fits.
}
\label{decaycurve}
\end{figure}

The experiments were done with samples fabricated by electron-beam lithography and shadow evaporation of Al films (Fig.~\ref{SEMetc}).
The qubit is the small superconducting loop intersected by four Josephson junctions \cite{Burkard05}, among which one is smaller than others by a factor of $\sim$0.56.
We have studied five similar samples, and here data from two representative samples are presented.
Spectroscopic measurements revealed that $I_p$ and $\Delta /h$ were 0.16~$\mu$A and  5.445~GHz for sample~A, and 0.34~$\mu$A and 5.08~GHz for sample~B, respectively.
The SQUIDs had maximum critical currents of $\sim$3.6~$\mu$A and $\sim$7~$\mu$A, and the switching currents under the flux bias used in the qubit readout were $\sim$1.6~$\mu$A and $\sim$4.6~$\mu$A, respectively.

The qubit is initialized to the ground state by relaxation, and is controlled by
a sequence of resonant microwave pulses.
The amplitude and the width of the pulses are adjusted so that the desired
qubit rotations are implemented.
At the end of the pulse sequence, the qubit state is read out by using the SQUID \cite{Chiorescu03}.
As the qubit in its eigenstates has a finite expectation value of the persistent current (except for $n_\phi = n_\phi^*$), critical current of the SQUID is slightly varied corresponding to each eigenstate.
Accordingly, the switching probability of the SQUID into the voltage state,
responding to a current pulse with a given amplitude and a
width, is different for each eigenstate of the qubit \cite{readout}.
The switching probability $P_{\rm sw}$, which thus reflects the population of the qubit eigenstates, is obtained by repeating the procedure --- initialization, control, and readout --- a number of times (typically $10^4$) and counting the switching events.

Figure~\ref{decaycurve} shows the results at the optimal bias condition $n_\phi = n_\phi^*$ and $I_b = I_b^*$, where both $\frac{\partial E_{01}}{\partial n_\phi}$ and $\frac{\partial n_\phi^*}{\partial I_b}$ were set to zero so that the dephasing due to fluctuations of $n_\phi$ and $I_b$ is minimal \cite{Bertet05,Burkard05}.
We find that energy-relaxation and echo signals decay exponentially with rates $\Gr$ and $\GdE $, respectively, and moreover $\GdE \simeq \Gr /2 $.
Therefore, at this bias condition the coherence observed in the echo measurement is mainly limited by the energy-relaxation process.
In the analysis of $P_{\rm sw}$ as a function of the readout current pulse height (data not shown), we have also found that the qubit mostly relaxes to the ground state, not to the mixture of the ground and excited states.
This excludes any classical noise from the possible source inducing the relaxation, and suggests that the spontaneous decay of the qubit is the dominant process.

\begin{figure}[b]
\includegraphics[width=75mm]{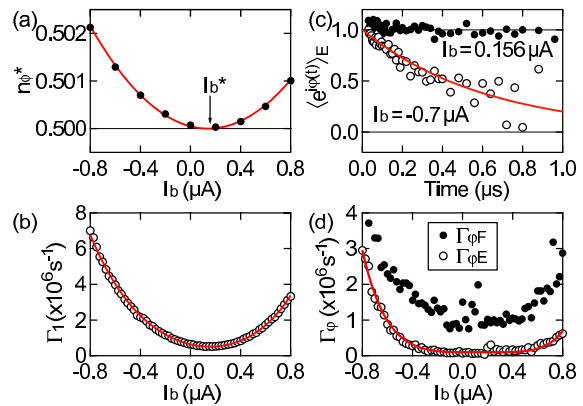}
\caption{
(color online).
Decoherence at various SQUID bias currents $I_b$ (sample~A).
Qubit flux bias is kept at $n_\phi = n_\phi^*$.
(a) Spectroscopically determined $n_\phi^*$ vs $I_b$.
Solid curve is a polynomial fit.
(b) Energy-relaxation rate $\Gr$ vs $I_b$.
(c) Dephasing component $\eiphiE$ in the echo measurements for different values of $I_b$.
Solid curve is an exponential decay, $\exp ( - \GpEexp\, t )$.
(d) FID dephasing rate $\GpFexp$, and echo dephasing rate $\GpEexp$ as a function of $I_b$.
See text for the explanations of the fitting curves in (b) and (d).
}
\label{Ib-dependence}
\end{figure}

Next, we study how the fluctuations of $I_b$ affect the coherence of the qubit.
We show that we can discriminate the contribution from that of $n_\phi$ fluctuations, even though the $I_b$ fluctuations also couple to the qubit through the flux degree of freedom $n_\phi^*$.
First, we present the result of qubit spectroscopy at various $I_b$.
For each $I_b$ we swept $n_\phi$ and found a minimum of $E_{01}$ to determine $n_\phi^*$ which is plotted in Fig.~\ref{Ib-dependence}(a).
Notice that at $n_\phi = n_\phi^*$, $E_{01}=\Delta$ and $\frac{\partial E_{01}}{\partial n_\phi} =0$.
The plot also represents how $I_b$ affects the flux in the qubit and how the fluctuations of $I_b$ are transformed into the fluctuations of $n_\phi^*$ via $\frac{\partial n_\phi^*}{\partial I_b}$.
The interpolated line $n_\phi^*(I_b)$ has a minimum at $I_b^* \equiv 0.156$~$\mu$A.
At $I_b = I_b^*$, the noise current coming through the SQUID bias line is split symmetrically into two branches of the SQUID and does not produce any fluctuations of the flux in the qubit, i.e., $\frac{\partial n_\phi^*}{\partial I_b} = 0$.
The small offset of $I_b^*$ from zero can be attributed to a small asymmetry in the SQUID junction parameters.

We find a few characteristic behaviors in the decay measurements at the bias conditions along the line $n_\phi = n_\phi^*(I_b)$ (Fig.~\ref{Ib-dependence}):
(i) The energy-relaxation and echo decay curves are always exponential for different values of $I_b$, while the rates $\Gr$ and $\GdE $ largely depend on $I_b$.
(ii) The ratio of pure dephasing rates in the echo measurement ($\GpEexp \equiv \GdE - \Gr /2 $) and the FID measurement ($\GpFexp \equiv \GdF - \Gr /2 $; with $\GdF$ obtained from exponential fitting of FID envelope \cite{commentFID}), $\GpEexp/\GpFexp$, is significantly smaller than 1 at $I_b\approx I_b^*$, while it approaches 1 at larger $|I_b-I_b^*|$.
These two facts, the exponential dephasing and the inefficient recovery of the coherence in echo measurement, suggest that the fluctuations introduced by $I_b$ have a relatively uniform spectrum at least up to $\GdE /(2\pi) \sim$10~MHz.
(iii) The $I_b$-dependent part of $\Gr$ and $\GpEexp$ are nicely fitted with a term proportional to $(\frac{\partial n_\phi^*}{\partial I_b} )^2$ and $(\frac{\partial n_\phi^*}{\partial I_b} )^4$, respectively.
This implies that the $I_b$-dependent contributions are solely due to $I_b$ fluctuations and that the fluctuations are well decoupled at $I_b^*$:
The $\Gr$ can be compared with the expectation from Fermi's golden rule at $n_\phi=n_\phi^*$, i.e., $\Gr = \frac{\pi}{2 \hbar^2} (\frac{\partial \varepsilon}{\partial n_\phi^*} )^2 (\frac{\partial n_\phi^*}{\partial I_b} )^2 S_{I_b}(\frac{E_{01}}{\hbar})$ \cite{Somega}.
Therefore, we can estimate the current noise spectrum at the qubit frequency $S_{I_b}(\frac{E_{01}}{\hbar}=2\pi \times 5.445~{\rm GHz})= 4.4 \times 10^{-27}$~A$^2$s/rad corresponding to the zero-point fluctuations of an impedance $\sim 270$~$\Omega$.
Also, the fourth power term in $\GpEexp$ can be understood as a result of the two-fold quadratic couplings between $E_{01}$ and $n_\phi^*$, and $n_\phi^*$ and $I_b$ \cite{Burkard05,Makhlin04}.

\begin{figure}
\includegraphics[width=75mm]{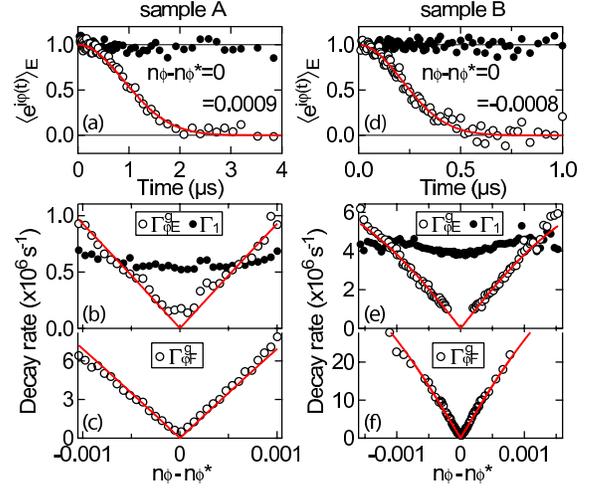}
\caption{
(color online).
Decoherence at various flux biases $n_\phi$.
SQUID bias current is kept at $I_b=I_b^*$.
(a) Dephasing component $\eiphiE$ in the echo measurements for different values of $n_\phi$ (sample A).
Solid curve is a Gaussian fit, $\exp [- (\GpE\, t)^2 ]$.
(b) Energy relaxation rate $\Gr$ and echo dephasing rate $\GpE$ vs $n_\phi$.
Solid line is a $|\frac{\partial E_{01}}{\partial n_\phi}| $ fit to $\GpE$.
(c) FID dephasing rate $\GpF$ vs $n_\phi$.
Solid line is a $|\frac{\partial E_{01}}{\partial n_\phi}| $ fit.
(d)-(f) Data for sample~B\@.
}
\label{gaussianfit}
\end{figure}

Having seen that the $I_b$ fluctuations can be well decoupled at $I_b= I_b^*$, we now focus on the effect of $n_\phi$ fluctuations.
Compared to the measurements above, measurements at $n_\phi \ne n_\phi^*$ show qualitatively different behaviors (Fig.~\ref{gaussianfit}):
(i) The energy-relaxation rate $\Gr$ is almost constant in the range of $n_\phi$, while the dephasing is strongly enhanced as $|n_\phi-n_\phi^*|$ is increased a little.
(ii) The echo decay curves are not purely exponential any more, and the extracted dephasing component $\eiphiE$ is fitted with a Gaussian decay, $\exp [ - (\GpE\, t)^2 ]$ \cite{deconvolution}.
(iii) The decay rate $\GpE$ increases almost linearly with $|n_\phi - n_\phi^*|$.

Because of the strong $n_\phi$-dependence of $\GpE$ at around $n_\phi^*$, we can rule out possible contributions of charge and critical-current fluctuations in the qubit.
Indeed, the observed behaviors are fully consistent with the presence of $1/f$ flux noise.
Let us assume Gaussian fluctuations of $n_\phi$ with the spectrum density $S_{n_\phi}(\omega) = \frac{A_{n_\phi}}{|\omega|}$.
Then, we calculate $\eiphiE = \exp [ -\frac{t^2}{2\hbar^2} (\frac{\partial E_{01}}{\partial n_\phi})^2 \int^\infty_{-\infty} \! d\omega \, S_{n_{\! \phi \!}} (\omega) \frac{\sin^4 (\omega t/4)}{(\omega t /4)^2} ] = \exp [ - (\GpE\, t)^2 ]$, where $\GpE = \frac{\sqrt{A_{n_\phi} \ln 2}}{\hbar} |\frac{\partial E_{01}}{\partial n_\phi}|  $ \cite{Cottet02}.
The Gaussian decay function is characteristic of the $1/f$ noise spectrum which has a singularity at low frequency.
The $n_\phi$-dependence of $\GpE$ follows $|\frac{\partial E_{01}}{\partial n_\phi}| = \frac{|\varepsilon|}{\sqrt{\varepsilon^2+\Delta^2}}$, which is basically linear in the range of $n_\phi$ studied, as shown in Figs.~4(b) and (e).

Free-induction decay measurement also supports the interpretation.
As shown in Figs.~\ref{gaussianfit}(c) and (f), the Gaussian decay rate $\GpF$ also increases almost linearly with $|n_\phi - n_\phi^*|$ \cite{commentFID}.
The asymptotic value of the ratio $\GpF / \GpE$ at large $|n_\phi - n_\phi^*|$ was between 4.5 and 7.5 for all the samples measured, which demonstrates that the echo procedure largely suppresses the dephasing and that the flux noise has a significant amount of low-frequency spectral weight.
The ratio $\GpF / \GpE$ was slightly larger than expected ($\sqrt{\ln \frac{1}{\omega_{ir} t} /\ln 2} \sim 5$ \cite{Ithier05}) for $1/f$ noise with an infrared cutoff frequency $\omega_{ir}/(2\pi) \sim 1$~Hz determined by the measurement procedure.
We could imagine there exist some additional low-frequency fluctuations contributing mainly to $\GpF$.
An alternative scenario is an ultraviolet cutoff which would make the ratio larger, as suggested in Ref.\cite{Ithier05} for $1/f$ charge noise.
However, such a cutoff should modify the Gaussian decay function of $\eiphiE$, which we did not observe.

We evaluated the amplitude of the $1/f$ flux noise from the $|\frac{\partial E_{01}}{\partial n_\phi}| $ fitting of $\GpE$, and obtained $\sqrt{A_{n_\phi}} = 9.3\times 10^{-7}$ and $9.9 \times 10^{-7} $  for samples A and B, respectively \cite{noise}.
Three other samples also gave $(1.1$-$2.0) \times 10^{-6} $ reproducively, though the origin of the $1/f$ flux noise remains to be understood.
One may suspect that the critical-current fluctuations of the SQUID junctions produce the $1/f$ flux noise via the change of the SQUID circulating current.
However, the normalized $1/f$ critical-current fluctuations of the reported value $\sim (10^{-6})^2$/Hz at 1~Hz \cite{Wellstood04,Eroms06} would give at most 1\% of the $\sqrt{A_{n_\phi}}$ for the mutual inductance of $\sim$20~pH between the qubit and the SQUID.
It may be worth mentioning that in SQUIDs with much larger dimensions similar amount of low-frequency flux noise has been known for a long time \cite{Wellstood87} and not yet understood either.


It is also interesting to compare the present result with a previous work on a flux qubit where the dominant dephasing was attributed to thermal fluctuations of photon numbers in the plasma mode of the readout SQUID \cite{Bertet05}.
There are a few noticeable differences in the result of Ref.\cite{Bertet05}:
(i) The echo decay was exponential.
(ii) $\GpEexp$ scaled with $( \frac{\partial E_{01}}{\partial n_\phi} )^2 $ as a function of $n_\phi$.
(iii) The echo decay rate was smallest at $n_\phi \ne n_\phi^*$ when $I_b \ne I_b^*$. In our experiment, $\GpE$ was smallest at $n_\phi = n_\phi^*$ for any $I_b$ (data not shown).
Indeed, our estimation shows that even in the presence of the same amount of $1/f$ flux noise as in our samples, the contribution would have been concealed by the photon-number fluctuations in the sample in Ref.\cite{Bertet05}.
In the present work, on the other hand, the dephasing rate due to the photon-number fluctuations is more than an order of magnitude smaller.
This is mainly because the coupling factor $| \frac{\partial^2 n_\phi^*}{\partial I_b^2}|$, which increases rapidly when the bias flux through the readout SQUID approaches half-integer flux quanta, is about five times smaller in our samples.


In conclusion, we have studied decoherence in flux qubits.
We presented qualitative difference between decay functions at different bias conditions and characterized the noises contributing to the decoherence.
We demonstrated that $1/f$ flux noise constituted one of the main decoherence sources.
On the other hand, the effect was suppressed at the optimal flux bias point where the energy relaxation process turned out to be the limiting factor of the coherence in the echo decay.
Reducing the energy relaxation rate is the key issue in improving the coherence of flux qubits further.


We are grateful to O. Astafiev, P. Bertet, J. Clarke, D. Est\`{e}ve, S. Lloyd, Yu.\ Makhlin, J. Martinis, and A. Shnirman for valuable discussions.


\begin{thebibliography}{99}
\bibitem{commentPhi}Here we approximately decompose the flux induced by the SQUID circulating current into the sum of an $I_b$-dependent part $\Phi_{I_b}$ and a part depending on the flux in the SQUID loop.
We define $\Phi_{\rm ex}$ including the latter term.
\bibitem{readout}The qubit state can be read out even if it is operated at $n_\phi =  n_\phi^*$ where the two eigenstates have zero expectation values of persistent current, since the applied current pulse adiabatically shifts the flux bias condition before reaching the switching current \cite{Chiorescu03}.
\bibitem{commentFID}Although FID curves at the optimal point often showed a beating-like envelope as seen in Fig.~\ref{decaycurve}(b), the curves were fitted better with a regular envelope when $|n_\phi - n_\phi^*|$ or $|I_b-I_b^*|$ is larger.
\bibitem{Somega}$S_\lambda (\omega) \equiv \frac{1}{2\pi} \int^\infty_{-\infty} d\tau \langle \widehat{\delta \lambda} (0) \widehat{\delta \lambda} (\tau) \rangle e^{-i\omega \tau}$ is defined as the unsymmetralized noise spectral density.
\bibitem{deconvolution}To deconvolute the decay curves and obtain $\GpE$, we fit the raw data $P_{\rm sw}(t)$ with $P_{\rm sw}^\infty + A \exp [ -\frac{\Gr}{2} t -(\GpE t)^2 ]$, using $\Gr$ obtained from energy relaxation measurement, and $P_{\rm sw}^\infty$, $A$, and $\GpE$ as free parameters in the least-square fit.
Then, $\eiphiE$ is calculated as $\frac{P_{\rm sw}(t) - P_{\rm sw}^\infty }{A \exp(-\Gr t/2)}$.
A similar procedure is used for FID.
\end{thebibliography}
\end{document}